\documentclass[%
 reprint,
 amsmath,amssymb,
 aps]{revtex4-1}
 
\usepackage{graphicx}
\usepackage[colorlinks=true,linkcolor=black, citecolor=black,
urlcolor=black]{hyperref} 

\usepackage{mathrsfs}
\usepackage[usenames,dvipsnames]{xcolor}
\usepackage{color}
\usepackage{graphicx}%
\usepackage{dcolumn}%
\usepackage{bm}%
\usepackage{tikz}
\usetikzlibrary{calc}

\newcommand{\bal}{\begin{align}}
\newcommand{\eal}{\end{align}}
\newcommand{\beq}{\begin{equation}}
\newcommand{\eeq}{\end{equation}}
\newcommand\beqa{\begin{eqnarray}}
\newcommand\eeqa{\end{eqnarray}}
\newcommand\bea{\begin{array}}
\newcommand\eea{\end{array}}
\newcommand\comment[1]{{}}

\newcommand{\sT}{{\mathscr T}}
\newcommand{\cT}{\CT}

    \newcommand{\COMMENT}[1]{}
    
    \newcommand{\neqa}{\nonumber\end{eqnarray}}

\def\a{{\alpha}}

\def\[{\left[}
\def\]{\right]}

\def\a{\alpha}

\def\[{\left[}
\def\]{\right]}

\def\<{\langle}
\def\>{\rangle}

\def\i2{\frac{i}{2}}

\def\bq{{\bar q}}

\def\be{\begin{eqnarray}}
\def\ee{\end{eqnarray}}
\def\no{\nonumber}

    \def\CK{{\cal K}}

    \def\CO{{\cal O}}

    \def\CT{{\cal T}}

    \def\<{\left\langle\,}
    \def\>{\, \right\rangle}
    \def\[{\left[}
    \def\]{\right]}

    \def\D{{\rm D}}

\def\Bm{B_{(-)}}
\def\Rm{R_{(-)}}
\def\Bp{B_{(+)}}
\def\Rp{R_{(+)}}

\newcommand{\Wpv}{{\hat {\sW}_{\rm pv}}}
\newcommand{\sW}{{W}}
\newcommand{\xh}{{\hat x}}
\newcommand{\hh}{{\hat h}}

\newcommand{\En}{{E}}

\begin{document}

\preprint{IPhT/t12/xxx}

\title{Six-loop Konishi anomalous dimension from the Y-system
  } 

\author{ S. Leurent$^{a,b}$, D. Serban$^{c}$, D. Volin$^{d}$}
  
\affiliation{%
\(^a\) Laboratoire de Physique The ́orique de l’ENS, 24, rue Lhomond,
75005 Paris, France\\
\(^b\) Theoretical Physics group, Imperial College, South
Kensington Campus, London SW7 2AZ, United Kingdom\\
\(^{c}\) Institut de Physique The ́orique, CNRS-URA 2306, C.E.A.-Saclay, F-91191 Gif-sur-Yvette, France\\
\(^{d}\) Nordita, KTH Royal Institute of Technology and Stockholm University, Roslagstullsbacken 23, SE-106 91 Stockholm, Sweden}

\begin{abstract}
We compute the Konishi anomalous dimension perturbatively up to six loop using the finite set of functional equations (FiNLIE) derived recently in 
\cite {GKLV}. The recursive procedure can be in principle extended to higher loops, the only obstacle being the complexity of the computation.  
\end{abstract}

 \maketitle
 
 \section{\label{sec:int}Introduction}

Using integrability in conjunction with the AdS/CFT correspondence led to some of the main achievements in theoretical high energy physics of the last decade.
One of them is determining the spectrum of the anomalous dimensions of
the planar \({\cal N}=4\) SUSY gauge theory, or equivalently finding 
the spectrum of free string theory in \(AdS_5\times S^5\) background. For a recent review on the subject see \cite{AdSReview}.  
For operators with large charges, the anomalous dimension is determined by a set of Bethe Ansatz equations, \cite{AdSReview}, while for operators with small charges, the information about the spectrum is encapsulated into an infinite set of functional equations derivable from the thermodynamic Bethe Ansatz (TBA)
equations 
\cite{Alyosha} and known under the name of \(Y\)-system. For the AdS/CFT integrable system, the \(Y\)-system was conjectured in
\cite{GKV} and derived from the the TBA equations in \cite{TBA:GKKV}
. The anomalous dimension of the Konishi operator, the shortest operator not protected by supersymmetry, is used as a testing ground for these equations, both for analytical and numerical computations. In particular, in perturbation four \cite{GKV} and five loop \cite{fiveTBA,fiveBH} corrections to the Bethe Ansatz were computed using  the \(Y\)-system, and they were found to coincide both with the corresponding L\"uscher  corrections \cite{JanikBaj,fiveL} and with the four-\cite{Kpert} and five-loop \cite{Eden:2012fe} perturbative gauge theory computations. At strong coupling, the result of the extrapolation to short, Konishi-like operators \cite{KS:GSSV}
agrees both with the numerical results \cite{GKVnum,Fnum} and with the string predictions 
\cite{KS:RT}.

Recently \cite{GKLV,BHfin}, the \(Y\)-system was reformulated in terms of a finite closed set of functional equations. In the present work, we  set up a recursive procedure to solve in perturbation the equations of \cite{GKLV} and we perform the explicit computation up to six loops. The procedure can be in principle continued to higher loops, in particular to the double-wrapping order at eight loops.
Double wrapping was already attained for the ground state energy in the twisted AdS/CFT
\cite{doublewrap}, however
our  case is significantly more complicated because of necessity to account the displacement of the Bethe roots.

\section{\label{sec:finlie}The set of functional equations}
  The wrapping corrections are encoded in a finite set of functional
  equations \cite{GKLV} which were derived from the AdS/CFT Y-system
  by solving the Hirota equation in the semi-infinite bands of the
  \(T\)-hook defined in \cite{GKV}.
The different solutions are  then glued together using analyticity constraints which reflect the physical properties the Y-system has to satisfy.
We give below a brief summary of this set of functional equations.
\paragraph{Input data.}  The functional equations depend on a specific operator through the position of the Bethe roots \(u_j\), via the objects
\(
B_{(\pm)}=\prod_{j=1}^M\sqrt{\frac{g}{\hat x_j^\mp}}\left(\frac{1}{x}-\hat x_j^\mp\right)\!,R_{(\pm)}=\prod_{j=1}^M\sqrt{\frac{g}{\hat x_j^\mp}}\left({x}-\hat x_j^\mp\right)\,\)
and the Baxter polynomial  \(Q^\pm=(-1)^MB_{(\pm)}R_{(\pm)}\).
Here we use the conventions \(f^{[a]}=f(u+ia/2)\), \(f^\pm=f^{[\pm1]}\), and  \(x(u)={u}/{2g}+i\sqrt{1-{u^2}/{4g^2}}\) is the Zhukovsky variable in the so-called mirror regime, with a cut on \(\check Z\equiv(-\infty,-2g]\cup[2g,\infty)\). For the Zhukovsky variable in the physical regime, with a branch cut on the interval \(\hat Z\equiv [-2g,2g]\),  we use the notation \(\hat x(u)={u}/{2g}+\sqrt{{u}/{2g}-1}\sqrt{u/2g+1}\). By convention, we denote with a hat the quantities which depend on \(\hat x(u)\), if we want to emphasize the position of the branch cut.
For the Konishi operator there are two magnons, \(M=2\), with \(u_1=-u_2=1/\sqrt{12}+\CO(g^2)\) and \(Q(u)=u^2-u_1^2\).
\paragraph{Parameterization of  the \(T\)- and \(q\)-functions.}
When considering a state from the \(sl(2)\) sector%
,
as it is the case for Konishi, 
the \(Y\)-system can be determined from only three functions: two real-valued densities with a finite support on \(\hat Z\), \(\rho\) and \(\rho_2\),  and  a complex-valued function \(U\) analytic in the upper half-plane.
A relatively simple formulation of the set of functional equations can be obtained using \(T\)-functions in two different gauges denoted \(\cT_{a,s}\) and \(\sT_{a,s}\).
The first gauge gives a simple solution of the Hirota equation in the right band \( s\geq a\)
\be
\label{cT}
\!\!\!\!\!\!\!\!\!\! \cT_{0,s}=1\;, \quad  \cT_{1,s}=s+\CK_s\ast \rho\;, \quad
  \cT_{2,s}=\hat \cT_{1,1}^{[+s]}\,\hat \cT_{1,1}^{[-s]}\;,
 \ee
with \((\CK\ast f)(u)=\frac{1}{2\pi i}\int_{-\infty}^{\infty}dv\,{f(v)}/({v-u})\)
and \(\CK_s\equiv\CK^{[s]}-\CK^{[-s]}\).
The second gauge gives a  solution of the Hirota equation in the upper band of the \(T\)-hook, \(a\geq|s|\),
\be
\label{curlyT}
\sT_{a,2}&=&q_{\emptyset}^{[+a]} \bq_\emptyset^{[-a]}\,,\  \quad \quad \sT_{a,-1}=(U^{[+a]}\bar
U^{[-a]})^2\sT_{a,1}\,,\\
\sT_{a,1}&=&q_1^{[+a]} \bq_2^{[-a]}+q_2^{[+a]} \bq_1^{[-a]}+q_3^{[+a]} \bq_4^{[-a]}+q_4^{[+a]} \bq_3^{[-a]}\,, \nonumber\\
\sT_{a,0}&=&q_{12}^{[+a]} \bq_{12}^{[-a]}+q_{34}^{[+a]} \bq_{34}^{[-a]} \nonumber \\
&-&q_{14}^{[+a]} \bq_{14}^{[-a]}-q_{23}^{[+a]} \bq_{23}^{[-a]}-q_{13}^{[+a]} \bq_{24}^{[-a]}-q_{24}^{[+a]} \bq_{13}^{[-a]}\,, \nonumber
\ee
We will also use the combinations \(\sT_{a,s}^c\) defined by (\ref{curlyT}) for \(a<|s|\).
The \(q\)-functions related among themselves by the Pl\"ucker relations \cite{GKLV}
are determined by
\(\rho_2\) and \(U\) as follows
 \begin{subequations}
\be
\label{q1q2}&&\!\!\!\!\!\!\!\!\!\!\!\!\!\!\!\!\!\!q_1=1\;, \ q_2=-iu+\CK*\rho_2-\CK\!*\!\Wpv\;,   \ \\
&&\!\!\!\!\!\!\!\!\!\!\!\! \!\!\!\!\!\!q_{12}=(u-u_1-\a)(u+u_1+\bar\a)\equiv Q+\delta q_{12}\,, \\
\label{eqqij}&&\!\!\!\!\!\!\!\!\!\!\!\!\!\!\!\!\!\!\frac{\{q_{13},q_{23}=q_{14},q_{24}\}}{q_{12}}=\sum_{k=0}^{\infty}\left[\frac{U^2\{1,q_2,q_2^2\}}{q_{12}^+q_{12}^-}\right]^{[2k+1]}\hspace{-2em},\\
&&\!\!\!\!\!\!\!\!\!\!\!\!\!\!\!\!\!\!q_{34}q_{12}=q_{13}q_{24}-q_{14}q_{23}\,,\ \  q_{\emptyset}q_{12}=q_2^--q_2^+\,,\\
&&\!\!\!\!\!\!\!\!\!\!\!\!\!\!\!\!\!\!q_3q_{12}^+=q_2q_{13}^+-q_{14}^+\,,\ \ q_{4}q_{12}^+=q_2q_{23}^+-q_{24}^+\,,\\
&&\!\!\!\!\!\!\!\!\!\!\!\!\!\!\!\!\!\!\sW_{a}=q_{3}^{[+a]}\bar q_{4}^{[-a]}+q_{4}^{[+a]}\bar q_{3}^{[-a]}\,,\
\ \ \sW\equiv\sW_0\,.
\ee
\end{subequations}
Above, \(\hat W_{\rm pv}=\frac{1}{2}(\hat W^{[+0]}+\hat W^{[-0]})\).  The definition
of  \(\rho_2(u)\) 
differs slightly from the one in \cite{GKLV},
so that here \(\rho_2(u)\) is of the square root type, in the sense
that \(\rho_2(u)/\sqrt{4g^2-u^2}\) is analytic in the vicinity of the
real axis.
 
\paragraph{Auxiliary integral equations.}
In the intermediate steps of the computations one needs to compute 3  quantities, \(Y_{1,1},Y_{2,2},\hat h\). \(Y_{1,1},Y_{2,2}\) are determined from the following relations, considered at \({\rm Im}(u)>0\): 
\begin{subequations}
\label{eqY}
\be
\label{eqYoY}
&&\!\ln\frac{Y_{1,1}\sT_{0,0}^-}{Y_{2,2}\sT_{1,0}}
\!\[\frac{\sT_{2,1}\cT_{1,1}^{-}}{\CT_{1,2}\sT_{1,1}^{-}}\]^2=2i\CK*{\rm
Im}\ln\left[\frac{\sT_{1,1}^{2}}{\sT_{0,0}\cT_{1,1}^{2}}\right]^{[1-0]}\hspace{-1em},\hspace{2em}\\
\label{eqYtY}&&\!\!\frac{1}{\hat{x}-{\hat{x}^{-1}}}\!\ln\frac1{Y_{1,1}Y_{2,2}}\frac{\sT_{1,0}}{\sT_{0,0}^-}\!=\!2\CK\!*\!{\rm Re}\!\left[\frac{1}{\hat{x}-{\hat{x}^{-1}}}\ln\frac{\sT_{1,0}}{\sT_{0,0}^+}\right]^{[-0]}\hspace{-1.5em}.
\ee
\end{subequations}
The function \(\hh\) is found from equations (5.38) and (6.14) in \cite{GKLV}.
 Only its large-volume asymptotic solution  (\ref{hathU}) is needed  for the six-loop computation.

\paragraph{Equations for \(\rho,\rho_2,U\).} After \(Y_{1,1},Y_{2,2},\hh\) are found, the set of functional equations can be closed by finding \(\rho,\rho_2\) from
\be
\label{magic}
\hspace{-1.5em}
\frac{1+Y_{2,2}}{1+Y_{1,1}^{-1}}&=&\frac{\sT^c_{1,2}\sT_{1,0}}{\sT_{1,1}^+\sT_{1,1}^-},\ 
\frac{1+Y_{2,2}^{-1}}{1+Y_{1,1}}=\frac{\hat\cT_{1,1}^{[1+0]}\hat\cT_{1,1}^{[-1-0]}}{\cT_{1,1}^+\cT_{1,1}^-}\;,
\ee
equations valid for \(u\in \hat Z\), and  determining \(U\)  from
\begin{equation}\label{eqU}
\hspace{-1.1em}
 \left[\frac{U}{\hh}
     \frac{\hh^{[2]}}{ U ^{[2]}}\right]^2=\frac{Y_{1,1}\sT_{0,0}^-}{Y_{2,2}\sT_{1,0}}
\!\[\frac{\sT_{2,1}\cT_{1,1}^{-}}{\CT_{1,2}\sT_{1,1}^{-}}\]^2   \[\frac{{Y_{1,1}Y_{2,2}}\sT_{0,0}^-}{\sT_{1,0}} \]^{[2]}    \hspace{-1em}
\end{equation}
for \({\rm Im}(u)>0\).
The solution of (\ref{magic}) has a one-parameter ambiguity fixed as explained below.
\paragraph{Supplementary constraints.}
Let us now emphasize the role of the Bethe roots. Bethe roots appear as the zeroes of the following functions:
\begin{subequations}
\be
\label{alphaconstraint}\hspace{-1.5em}\sT_{1,0}^+(u_j)=0\,,&&{\rm \ fixes\  the\ value\ of}\   \a\,,\\
\label{T11constraint}\hspace{-1.5em}\sT_{1,1}(u_j)=0\,,&&{\rm \ fixes\  the\ ambiguity\ in \ \eqref{magic}  }.
\ee
Using them and the Hirota equation \(\sT_{1,0}^+\sT_{1,0}^-=\sT_{0,0}\sT_{2,0}+(U^+\bar U^-)^2\sT_{1,1}^2\), we conclude that \(\sT_{0,0}\sT_{2,0}\) should have a double zero at Bethe root which appears to be a double zero of \(\sT_{0,0}\). 
\end{subequations}
The overall normalization of \(U\) is not fixed by \eqref{eqU} and should be defined from
\be\label{Unorm}
U\bar U=\sqrt{\sT_{0,0}^+\sT_{0,0}^-}\frac{1-Y_{1,1}Y_{2,2}}{\sT_{0,1}^c}\,\ \ u\in \hat Z\,.
\ee
\paragraph{Exact Bethe equations.}
The set of equations above has a solution for a range of values  of \(u_1\). To get the correct answer for the energy, one has to insert the value  of \(u_1\) which is fixed by the exact Bethe equation, which can be written  \cite{GKLV} in the form
\begin{equation}
   -\left[\frac{\hat h^-}{\hat
       h^+}\right]^2\frac{Y_{2,2}^+}{Y_{2,2}^-}\frac{\CT_{1,2}^+}{\CT_{1,2}^-}\frac{\hat\CT_{1,1}^{[-2]}}{\hat\CT_{1,1}^{[+2]}}=1\ \ {\rm at}\ \ u=u_j\,.
\label{eq:BetheAn}
\end{equation}

\noindent {\bf Asymptotic solution.} The equations listed above depend on the parameter \(L\) which sets the large \(u\) behavior of  \(\hh\). In the large volume limit \(L\to\infty\) the Y-system, and hence the functional equations above, can be solved explicitly \cite{GKV}. All the functions \(q_{ij}\), with the exception of \(q_{12}\), and \(q_3\) and \(q_4\) are suppressed at least by a factor \(\xh^{-L}\) with respect to \(q_1\), \(q_2\) and \(q_{12}\), so they are zero in the asymptotic expressions of \(\sT_{a,s}\). \(q_2\) and \(q_{12}\) are  in this limit given by
\(
(q_2)_{\rm as}=-iu+\CK*\rho_2\;, 
 \
 (q_{12})_{\rm as}=Q\,.\)
The asymptotic values of \(\rho\) and \(\rho_2\) are
\be
\hspace{-1em} (\rho)_{\rm as}=4\,\frac{\sqrt{4g^2-u^2}}{\En_{\rm as}+2},\
(\rho_2)_{\rm as}\!=-4\,\frac{\sqrt{4g^2-u^2}}{{\En_{\rm as}}-2},
\ee
where \(E_{\rm as}\) is defined in (\ref{wrapen}). 
One can check that both the equations  (\ref{eqY}) and (\ref{magic}) are satisfied by
\(\left(Y_{1,1}Y_{2,2}\right)_{\rm as}\!\!=\!\frac{\Bm\Rp}{\Bp\Rm}\) and \(\left(\frac{Y_{1,1}}{Y_{2,2}}\frac{\sT_{2,1}^2}{\cT_{1,2}^2}\right)_{\rm
as}\!\!\!=\left(\!\frac{\En_{\rm as}+2}{\En_{\rm as}-2}\right)^2\frac {Q^+}{Q^-}\frac{\Bp^{[-2]}x^{[-2]}}{\Bm^{[+2]}x^{[+2]}}\).
The asymptotic values of \(\hh\) and \(\hat U\) are given by
\be
\hspace{-1.5em}\frac{{\hat x^{\frac L2+1}}(\hh)_{\rm as}}{\Lambda_h}\!=\!
\frac{{\xh^{\frac L2}}(\hat U)_{\rm as}}{\Lambda_U{\hat B_{(-)}}}\!\!\left(\frac{\hat B_{(+)}}{\hat B_{(-)}}\right)^{\!\!\!\frac{\D^2}{1-\D^2}}\!\!\!\!=
\prod_{j=1}^2\frac{e^{i\chi(\xh,\xh_j^- )}}{e^{i\chi(\xh,\xh_j^+)}},
\label{hathU}
\ee
where for \(\chi(x,y)\) one can use the BES perturbative expansion 
\cite{BES}. The shift operator \(\D\) is defined such that \(\D
f=f^D=f^+\).
The overall normalization \(\Lambda_h\) is irrelevant, {\it cf.} \eqref{eq:BetheAn} and \eqref{eqU}, whereas \(\Lambda_{U}=\frac 12\sqrt{E_{\rm as}(E_{\rm as}-2)}\) 
\(\times \exp({6\gamma\,
  g^2(-1\!+\!4g^2-28g^4)\!+\!18\zeta_3g^4-\!24\,g^6(3\zeta_3+5\zeta_5)})\) is
fixed by \eqref{Unorm}. The constant \(\gamma\) depends on the regularization scheme for diverging sum. We use the prescription
\(\frac{1}{1-\D^2}\frac 1{u}=-i\,\psi(-iu)\)  for which \(\gamma\)  is the Euler constant.

Although for Konishi-like operators \(L\) is  not large, the asymptotic solution is still valid up to  at least \(L\) loops (four loops for Konishi), so the weak coupling is effectively a large volume limit.  We use as a constraint that the full solution should reduce to the asymptotic one  at weak coupling.
In the following, we will  continue to call  `asymptotic'  the above
quantities evaluated at the {\it exact} position of the Bethe roots
\(u_j\), determined by equations (\ref{eq:BetheAn}). These quantities
will therefore incorporate part of the wrapping corrections via the
corrections to the Bethe roots.

\section{\label{sec:Weak}Weak coupling expansion}

In order to solve perturbatively the functional equations, our strategy is to subtract from the exact equations the asymptotic ones and to use the fact that the deviations of the \(T\) functions from the asymptotic values are small. The resulting equations depend on the ratio of the exact and the asymptotic values, so we denote \((T)_{\rm r}\equiv T/(T)_{\rm as}\). The following quantities enter the functional equations
\be
\hspace{-1.5em}
H\!=\!\ln \!
\left(\frac{\sT_{1,0}}{\sT_{0,0}^+}\right)_{\!\!\rm r}\!,\; r\!=\!\ln\! \left(\frac{\sT_{1,1}}{\CT_{1,1}}\right)_{\!\!\rm r}\!, \;
\hat r_*\!=\ln \! \left(\frac{\hat q_2^+- \hat q_2^{-}}{\hat\CT_{1,1}}\right)_{\!\!\rm r}\hspace{-.2em}.
\ee
All these quantities are small in perturbation, and this will allow performing the expansion of the functional equations.  
For example, up to seven loops, one has
\be\label{Happr}
        H\simeq\frac{{\delta \bar q_{12}}^-}{Q^-}-\frac{{\delta \bar q_{12}}^+}{Q^+}+\frac{(q_{24}^++2q_{14}^+\bar{
q}_2+q_{13}^+\bar q_2^2)\bar U^2}{(Q^+)^2\,Q^-}. \ \ \ \
\ee
The quantities \(U\) and \(q_2\) are determined asymptotically from (\ref{q1q2}) and (\ref{hathU}), and at the leading order they are equal to
\be 
U^2=-\frac{2g^4}{u^2}+\ldots,\ q_2=-i\,u-\frac i{3}\frac 1u+\ldots\,.
\ee
As for the \(q\)-functions \(q_{13}, q_{23}=q_{14}, q_{24}\), they are given by the sums in (\ref{eqqij}) and in perturbation they have an array 
of equidistant poles which will be responsible for the appearance of the zeta functions in the final answer. 
For example, at the leading order, we have
\be
  q_{13}=&18\,g^{4}\left(-i\,u+Q \,\,\psi^{(1)}(-i\,u+1/2)\right)\,,
\ee
where \(\psi^{(n)}\) denotes the  \(n\)-th derivative of the digamma function. %
The functions \(q_3\) and \(q_4\) have a similar structure. \(q_{34}\)
is given by a double sum and it contributes only at eight loops and
higher. A priori, the infinite sum in (\ref{eqqij}) creates also
infinitely many poles at  positions of the shifted Bethe
roots. However, because of the equality  \((\hat U^+/\hat
U^-)^2=-Q^{[+2]}/Q^{[-2]}\) which holds at the Bethe root at first
four nontrivial orders and which is just the asymptotic Bethe
equation, the poles in \(q_{13}\) cancel out pairwise (except for the
first one, which is cancelled by an overall factor \(Q\)). 
The cancellation mechanism still holds for \(q_{14}\) and \(q_{24}\)
because  \((\hat q_2^++\hat{\bar q}_2^-)_{\rm as}=(\hat\sT_{1,1})_{\rm
  as}\) and \(\hat\sT_{1,1}(u_1)=0\). 
At least up to seven loops, the \(q\)-functions 
are given by linear combination of the \emph{multiple Hurwitz zeta functions,}
which we define by
\begin{eqnarray}
\label{MHZ}
     \!\! \eta_{a_1,a_2,\cdots,a_n}(u)&=&\!\sum_{k=0}^{\infty}\left(\frac
    1 {u+i k}\right)^{a_1}\!\!\! \!\eta_{a_2,\cdots,a_n}\!(u+i(k+1)),\no\\
     \eta_a(u)&=&\,\frac{i^a}{(a-1)!}\;\psi^{(a-1)}(-i\,u)\;,\ \ a\geq 1,
\end{eqnarray}
with coefficients which are rational functions of \(u\) .

\begin{figure}
  \centering
\includegraphics{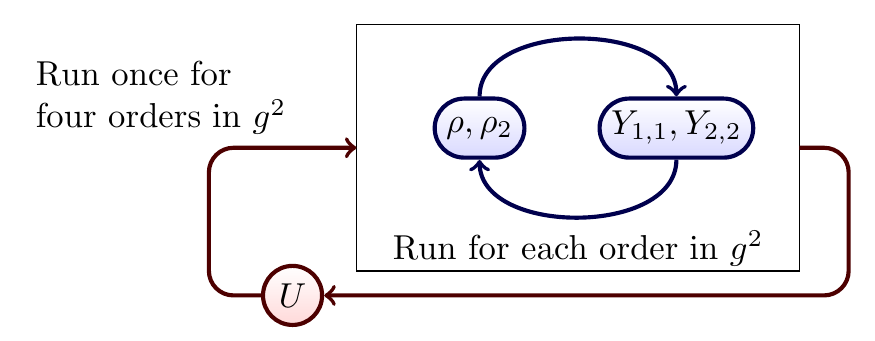}
\vspace{-.5cm}
  \caption{\label{fig:iterative}Structure of the perturbative computation.}
\end{figure}
The algorithm for computing the perturbative computation is summarized
in figure \ref{fig:iterative}. The interior loop determines the
densities \(\rho\) and \(\rho_2\), or rather their variations with respect to the asymptotic values, \(\delta\rho\) and
\(\delta\rho_2\), from equations (\ref{eqY}), analytically continued to \(u\in\hat Z\), and from (\ref{magic}). 
In order to preserve the square root structure of the 
two densities, we need to make a rescaling \(z=u/2g\). When performing various integrations, we encounter two different situations.
In the first, the integration is (parallel to) the real axis and we deform the contour to
pass  just below it and to avoid the possible singularities on the real axis.
This allows us to perform uniform expansion in \(g\). In the second situation the Cauchy kernel \(\CK\) acts on a function with a finite support, like \(\delta\rho\) and \(\delta\rho_2\), then it can be  expanded in terms of moments:
\be\label{momenta}
      \CK^{[s]}\ast\rho =\sum_{n\geq 0}\frac{i (2g)^{2n+1}}{(u+is/2)^{2n+1}}\int_{-1}^1\frac {dz}{2\pi}z^{2n}\,\rho(z),
\ee
 except for \(s=0\), where
\(
  \CK^{[\pm 0]}\ast\rho=\pm\rho/2\ +\ \slash\!\!\!\CK\ast\rho\,,
\)
the slash meaning principal value.
The equations (\ref{eqY}) and (\ref{magic}) finally reduce to the linear system
\be\label{leadrhorho2}
\left(\begin{array}{cc}
2 & 12^{} \\
1 & 3 \\
\end{array}\right)
\left(\begin{array}{c}
v_1 \\
g^{2 }v_2 \\
\end{array}\right)=g^9\left(\begin{array}{c}
C_1 \\
C_2 \\
\end{array}\right),
\ee
where \(C_1\) and \(C_2\) are Taylor series in \(g^2\) and \(z\)  is known from the previous orders of perturbative expansion (at the leading order in \(g\) they are  constants) and
\be
v_1=\frac{\,\delta\rho}{\sqrt{1-z^2}},\ v_2=\!\left(\frac{(1-2z^2)\delta\rho_2}{\sqrt{1-z^2}}-4iz\ \slash\!\!\! \CK\!\ast\!\delta\rho_2\right)\!\!.\ \ \
\ee
By definition of \(\rho\) (and \(\rho_2\)), we know that
\(\rho(2gz)/\sqrt{1-z^2}\) is analytic in the vicinity of the real
axis and hence can be Taylor-expanded. At a given order in \(g\), this means
that \(\delta \rho\) and \(\delta\rho_2\) are polynomial in
\(z\) times \(\sqrt{1-z^2}\). Using this information, the equation  (\ref{leadrhorho2}) gives, at leading order,
\be
\hspace{-1em}\delta\rho_2=g^7(A+Bz^2)\sqrt{1-z^2}\,,\ \ \delta\rho=g^9\,C\sqrt{1-z^2}\,,\ \  
\ee
with \(6A=C_1-2C_2\), \(2C=-C_1+4C_2\). The value of \(B\) is unconstrained
by (\ref{leadrhorho2}), and it is fixed by 
\eqref{T11constraint}.

 \vspace{1em}
{\bf The result for the energy.}
The energy can be computed \cite{GKLV} from the behavior at  large \(u\) of
\(\ln Y_{1,1}Y_{2,2}\),
\(\ln Y_{1,1}Y_{2,2}\simeq {iE}/{u}\).
Isolating the asymptotic and the wrapping part in the above expression, \(E=
E_{\rm as}+E_{\rm wrap}\), one obtains
\be 
\hspace{-2em}
E_{\rm as}=2-8\,g\,{\rm Im}\left(\frac 1{\hat x_1^+}\right)\!,\;
\label{wrapen}
        E_{\rm wrap}=\int_{\mathbb{R}-i0}\frac {-H(u)du}{\pi\sqrt{1-\frac{4g^2}{u^2}}}\,.
\ee
The Asymptotic Bethe Ansatz \cite{BES} predicts  up to 6 loops: 
\be
&&E_{\rm BAE}=E_{\rm as}(u_1\to u_{1,\rm BAE})=2+12g^2-48g^4+336\,g^6\no\\
&&-(2820+288\,\zeta_3)g^8+(26508+4320\,\zeta_3+2880\,\zeta_5)\,g^{10}\no\\
&&-(269148+55296\,\zeta_3+44064\,\zeta_5+30240\,\zeta_7)\,g^{12}\,.
\ee
At finite volume (\(L=2\) for the Konishi operator), the energy receives
corrections both from \(\En_{\rm as}\) through the correction of the position of Bethe roots \(u_{1}\), and from \(E_{\rm wrap}\).
The corrections to the  Bethe equations leading to displacement of the Bethe roots start at \(5\) loops, where  they are due to correction of \(Y_{2,2}\) only. At 6 loops one should take into account the first corrections to \(\rho_2\), whereas effects from correction to \(\rho\) and \(\hh\) are delayed at least up to 7 loops. 
\(E_{\rm wrap}\) is non-zero starting from 4 loops. 
The single-wrapping corrections, up to seven loop, follow from computations within interior loop in 
figure \ref{fig:iterative}, whereas exterior loop has to be run only once, to find the explicit analytic expression (\ref{hathU}). 
For   \(H\) one can use the approximation (\ref{Happr}). The double wrapping effects, in particular correction to \((U)_{\rm as}\),  are important starting from 8 loops.

We have performed the explicit perturbative expansions discussed in
previous sections and  computed \(\delta E\equiv E-E_{\rm BAE}\) up to six loops,
i.e. up to \(g^{12}\) term. Intermediate expressions are too bulky to be
presented  here. We summarized them in the {\it Mathematica} notebook
file  \cite{link}. They 
contain \(\eta\)-functions (\ref{MHZ}) and their residues at  the Bethe root. However,
the final expression is significantly simpler and is given in terms of zeta-functions:\be
&&\delta\En_{4\&5\ \rm loop}=(324 + 864\,  \zeta_{3} - 1440\, \zeta_{5})g^8+\\
&&(-11340+2592\, \zeta _3-11520\,
   \zeta _5 -5184\, \zeta _3^2+30240\, \zeta _7)g^{10},\no\\ 
&&\delta\En_{6\ \rm loop}=(261468 - {207360} \zeta_3 - {20736}
\zeta_3^2 +  156384 \zeta_5 \no\\
&& \hspace{2em}
+  {155520} \zeta_3 \zeta_5 + 
 105840 \zeta_7 - 489888 \zeta_9 )g^{12} \,. 
\ee
At four and five loops we reproduced the already known answers
\cite{JanikBaj,fiveL}.
Our final expression for the Energy of the Konishi operator is
\begin{align}
  E=&2 + 12 g^2 - 48 g^4 + 336 g^6 + 
(-2496 + 576 \zeta_3 \nonumber\\-& 1440 \zeta_5)  g^8  + 
 (15168 + 6912 \zeta_3 - 5184 \zeta_3^2 - 8640 \zeta_5 \nonumber\\+& 
    30240 \zeta_7)  g^{10}+ 
 (-7680 - 262656 \zeta_3 - 20736 \zeta_3^2 \nonumber\\+& 112320 \zeta_5 + 
    155520 \zeta_3 \zeta_5 + 75600 \zeta_7 - 489888 \zeta_9) g^{12}
\end{align}
 We were informed that  Z.~Bajnok and R.~Janik
have obtained \cite{BJ6loop} six and seven-loop corrections using
L\"uscher's method; our result coincides with their six loop
result. Also, fitting the known numerical results
\cite{GKVnum,Fnum,Gromovprivate} with a diagonal Pad\'e approximant,
we were able to fix the 6-loop energy with \(5\%\) confidence. Our
analytic result is compatible with this numerical estimation.

 \section{Conclusion} 
 
We have computed the wrapping corrections of the Konishi operator up
to  6-loop order using the functional equations proposed in
\cite{GKLV}. We have adjusted the structure of the functional
equations for a systematic perturbative expansion and   
 we hope to be able to apply our methods to reach double-wrapping
 orders for the Konishi states.    An interesting question to explore
 is whether the cancellation of poles at Bethe roots observed when
 computing \(q_{ij}\)  holds at any order and if it can be used as a
 regularity condition implying the exact Bethe equation.  Another
 question is what type of functions  appear in the final answer. So
 far the expression for the energy is reducing to  Euler-Zagier sums
 and we believe that it will be always so. 
 \vspace{1em}
  \begin{acknowledgments}
   The authors are indebted to  Z. Bajnok, S. Frolov,  V. Kazakov,
   I. Kostov   and  especially to N. Gromov for useful discussions and
   to I. Shenderovich and E. Sobko for involvement in the early stages
   of this project. 
   D.S. and S.L. thank Nordita for hospitality, where  part of this
   work was done. 
\end{acknowledgments}

\providecommand{\href}[2]{#2}\begingroup\raggedright

\endgroup

\end{document}